\documentclass{article}
\usepackage[latin1]{inputenc}
\usepackage{amsmath}
\usepackage{setspace}
\newfont{\mcal}{rsfs10 scaled 1200}
\usepackage{latexsym}
\begin{document}

\hfill{}TUW--99--14

{\par\centering {\huge Hawking radiation from dilaton gravity in
1 + 1 dimensions: a pedagogical review}\huge \par}
\bigskip{}

{\par\centering {\Large W. Kummer}\footnote{%
\( ^{\star } \)email:wkummer@tph.tuwien.ac.at
}\( ^{\star } \)\par}
\smallskip{}

{\par\centering {\small and }\small \par}

{\par\centering {\Large D.V. Vassilevich}\footnote{%
\( ^{\dagger } \)e-mail: Dmitri.Vassilevich@itp.uni-leipzig.de

\( ^{\ddagger } \)On leave from Department of 
Theoretical Physics, St. Petersburg
University, 198904 St. Petersburg, Russia
} \( ^{\dagger \ddagger } \)\par}

\bigskip{}
{\par\centering \( ^{1} \)Institut f\"{u}r 
Theoretische Physik, Technische Universit\"{a}t Wien\\
Wiedner Hauptstr. 8-10, A-1040 Wien\\
Austria{\large }\\
 {\large \( ^{2} \)} Institut 
f\"{u}r Theoretische Physik, Universit\"{a}t Leipzig,\\
Augustusplatz 10, D-04109 Leipzig,\\
Germany\par}

\begin{abstract}
Hawking radiation in \( d=4 \) is regarded as a well understood
quantum theoretical feature of Black Holes or of other geometric
backgrounds with an event horizon.  On the other hand, the
dilaton theory emerging after spherical reduction and
generalized dilaton theories only during the last years became
the subject of numerous studies which unveiled a surprisingly
difficult situation.  Recently we have found some solution to
the problem of Hawking flux in spherically reduced gravity which
has the merit of using a minimal input.  It leads to exact
cancellation of negative contributions to this radiative flux,
encountered in other approaches at infinity, so that our result
asymptotically coincides with the one of minimally coupled
scalars.  The use of an integrated action is avoided - although
we have been able to present also that quantity in a closed
expression.  This short review also summarizes and critically
discusses recent activities in this field, including the problem
of ``conformal frames'' for the background and questions which
seem to be open in our own approach as well as in others. 
\end{abstract} 

\section{Introduction}

The last years have seen an increased interest in dilaton
theories, treated in \( 1+1 \) dimensions.  The main motivation
to study such theories derives from the fact that spherical
reduction of \( D \)-dimensional Einstein gravity (SRG)
precisely generates a theory of this type \cite{Thom 84}.

Consider a line element
\begin{equation}
\label{1}
(ds)^{2}=g_{\mu \nu }dx^{\mu }dx^{\nu }-\frac{4}{\lambda
^{2}}e^{-\frac{4}{D-2}\phi }(d\Omega )^{2},
\end{equation}
 where \( d\Omega  \) is the standard surface element on the 
 \( D-2 \)-dimensional unit 
sphere \( S^{D-2} \). The dilaton field \( \phi  \) depends on the two first
coordinates \( x^{\mu } \) only. Then the \( D \)-dimensional Einstein-Hilbert
action, after \( \int d^2\Omega  \) has been dropped, becomes the one of SRG
\begin{equation}
\label{2}
{\mathcal{L}}_{SRG}=e^{-2\phi }\sqrt{-g}\left( R+\frac{4(D-3)}{D-2}(\nabla \phi )^{2}-\frac{\lambda ^{2}}{4}(D-2)(D-3)e^{\frac{4}{d-2}\phi }\right) ,
\end{equation}
which represents a particular dilaton theory with the Schwarzschild
Black hole (SBH) in \( D \) dimensions as its general classical solution.

As of 1991 also the special case \( D\rightarrow \infty \), \(
\lambda ^{2}\, D^{2}\rightarrow const.  \) (dilaton Black Hole,
DBH) received particular attention \cite{Man 91}.  Inspired by
string theory \cite{Eli 91} this model also exhibits a BH
solution, although its singularity is null-complete \cite{Kat
97}.  On the other hand, (\ref{2}) allows a (classical) solution
even when coupling to matter is introduced.

However, until quite recently in the latter case a quantum
treatment of an action like (\ref{2}) which went beyond the
semiclassical calculations of the original literature was not
known.  In several recent papers \cite{KLV 97-99} the authors
have shown (together with H.  Liebl) that a quantum treatment
including full back-reaction for \( 2D \) gravity can be
formulated in which ``geometry'' is integrated out (trivially)
and matter can be considered in a loop-wise expansion. 
Intriguing effects of ``BH production'' can be observed
\cite{KV99}.

A necessary prerequisite for the plausibility of \( 2D \)
quantum gravity arguments of such a type is that a simpler
consequence of Einstein theory in \( D=4 \) is consistently
reproduced at the level of a SRG action like (\ref{2}) or its
generalizations.  This simpler consequence is Hawking radiation
from a \textit{fixed} SBH background which is determined by the
solution of (\ref{2}).  For a large BH (\( m_{BH}>>m_{Planck}
\)) this should be an excellent approximation.  Of course, it
neglects quantum effects from gravity.

Somewhat surprisingly until the seminal work of Mukhanov, Wipf
and Zelnikov \cite{Muk 94} no calculation to that effect had
been performed - probably because no differences had been
expected with respect to the well-known \( D=4 \) approach
\cite{Haw 75,UN 76,Chris 77}.  It was noted by the authors of
\cite{Muk 94} that technical difficulties arise, when the matter
Lagrangian with minimal coupling of massless scalars \( f \) in
\( D=4 \) (\( a=0,1,2,3 \))
\begin{equation}
\label{3}
^{(4)}\mathcal{L}=\frac{1}{2}\sqrt{-^{(4)}g}\,
^{(4)}g^{ab}\left( \partial _{a}f\right) \left( \partial
_{b}f\right)
\end{equation}
is spherically reduced. In its \( 2D \) form a dependence on the
dilaton field \( \phi  \) appears
\begin{equation}
\label{4}
{\mathcal{L}}^{(nm)}=\frac{1}{2}e^{-2\phi }\sqrt{-g}g^{\mu \nu }(\partial _{\mu }f)(\partial _{\nu }f)
\end{equation}
 which must be taken into account carefully. Otherwise inevitably a negative
flux of energy at infinity is related to the radiating BH.

The problem resurfaced again in 1997 as a consequence of a paper
by R.  Bousso and S.W.  Hawking \cite{BH 97/1} which
contradicted \cite{Muk 94} and work of the present authors
together with H.  Liebl \cite{KLV 97/H}, as well as others
\cite{ICH/97}.  In \cite{KLV 97/H} the problem had been
considered for a generalization of (\ref{2}) to the action
\cite{Kat 97}
\begin{equation}
\label{5}
L=\int d^{2}x\sqrt{-g}e^{-2\phi }(R+4a(\nabla \phi
)^{2}+Be^{2(1-a-b)\phi }).
\end{equation}
of which (\ref{2}) is the particular case \( a=\frac{D-3}{D-2} \), \( b=a-1 \)
and \( B=-\frac{\lambda ^{2}}{4}(D-2)(D-3) \). 

This action not only covers SRG, but also the DBH \cite{Man 91}.
For \( a=1 \), \( b=0 \), the Jackiw-Teitelboim model \cite{jackiw} \( a=0 \),
\( b=1 \) is included. Mignemi \cite{mignemi} considers \( a=1 \) and all
values of \( b \). The models of ref.\ \cite{fabbri} correspond to \( b=0,\, 
a\leq 1 \). 
Lemos and Sa \cite{lemos} discussed global solutions for \( b=1-a \) and all
values of \( a \).

In \cite{Kat 97} the global properties of all models of this
type were analyzed in a systematic way (cf.  the table Fig.  6
in the second ref.\  of \cite{Kat 97}).  The action (5) includes
all \( 2D \) theories with one singularity and one horizon. 
Only along the line \( b=1-a \) in the interval \( 0<a<1 \) (or
\( 3<D<\infty \)) the conditions of a ``Minkowski ground state''
theory \cite{Kat 97} are fulfilled, i.e.  the presence of the
singularity is tied to a non-vanishing Arnowitt-Deser-Misner
(ADM) mass \cite{ADM} which
in \( D=2 \) is proportional to an absolutely (in space and
time) conserved quantity, even when general matter interactions
are included \cite{KS92}.  Also only for those models the
conformal (Carter - Penrose) diagram coincides with the one of
SRG.

Already before the study of Hawking radiation for nonminimally
coupled scalars as in (\ref{4}), where \( \phi \) was even
generalized to a function $\varphi ( \phi )$, minimally coupled
scalars had been considered on such backgrounds in this context
\cite{LVA}.  Here - even for asymptotic de Sitter and Rindler
space-times - no problems with the flux had arisen.  But, as
pointed out in \cite{KLV 97/H} for nonminimal coupling a
``straightforward'' application of arguments, valid in \( d=4 \),
to \( D=2 \) resulted in a negative flux at infinity - even when
some mistaken conceptions regarding the conformal anomaly in
\cite{BH 97/1} were corrected \cite{KLV 98/H}.

In \( D=4 \) two standard lines of arguments are used to derive
the flux of Hawking radiation to infinity and its relation to
Hawking-temperature, the temperature at the horizon seen by an
asymptotic observer (we restrict ourselves to asymptotically
flat situations, for simplicity).

The first - and most well-known - one has as its origin in a
specific application of the Unruh effect \cite{UN,UN 76}: For
two systems in relative accelerated motion the quantum vacuum of
one system appears as a state with thermally distributed
particles of a temperature which is proportional to the
acceleration.  Surface gravity, a geometric quantity determined
by the normal derivative of the Killing norm at the
(nondegenerate) horizon, in this way turns out to be
proportional to the Hawking temperature \cite{GH 76}.  On the
other hand, the flux of radiation to infinity is computed from
Bogoliubov coefficients which - during the the formation of a BH
- relate incoming spherical waves before the collapse with
outgoing ones from the neighbourhood of the horizon \cite{Haw
75,UN 76}.  This flux shows BH radiation with the BH behaving
like a black body at the Hawking temperature.

Very soon it was realized that the formation of the BH is of
relatively little importance for the bulk effect.  What is
really relevant are the null directions from the horizon to
light-like infinity (\( \mathcal{I}^{+} \)).  Therefore, the
second main approach due to Christensen and Fulling (CF)
\cite{Chris 77} restricts itself mostly to that region. 
Integrating the conservation law of the energy momentum flux to
infinity for minimally coupled scalars in \( D=2 \) the correct
Hawking flux was obtained.  Here the ``quantum input'' is the
one loop quantum correction to the trace of the energy momentum
(EM) tensor derived from (\ref{3})\footnote{
a traceless EM tensor is necessary at the classical level.  Such
a tensor does not follow from (\ref{3}), but requires there an
additional term \( -R\sqrt{-g}f^{2} \).  For the Ricci-flat
background of a BH such a term vanishes.  } and the assumption
that the radiation flux should stay finite at the horizon (Unruh
vacuum \cite{UN 76}), when considered in terms of global
(Kruskal - Szekeres) coordinates.  However, in the computation
in \( D=4 \) several functions remain undetermined, even when
all quantities are assumed to depend on time and one radial
coordinate only.

In a calculation at the spherically reduced level (\( D=2 \),
eq.\ (\ref{5})) clearly both these approaches should be viable -
at least in principle.  In addition, choosing the conformal
gauge \( g_{\mu \nu }=\eta _{\mu \nu } \exp \, \left( 2\rho
\right) \) the dependence on a single function \( \rho \) allows
the functional integration of the trace anomaly to an effective
action.  The dependence of \( \rho \) can be ``covariantized''
to the (nonlocal) Polyakov action \cite{POL 81} for scalars
coupled minimally in \( D=2 \) (i.e.  without the factor \( \exp
\, \left( -2\phi \right) \) in (\ref{4})) or for some
generalization thereof, if an additional dependence on the
dilaton field exists.  From that effective action the EM tensor
and, as a consequence, also the flux to (asymptotically flat)
infinity should be calculable.  This line had been followed in
ref.\  \cite{Muk 94}, where the authors realized that another
non-Weyl-invariant piece must be added to the action which would
be missed in a naive functional integration.  However that piece
could not be obtained exactly.  Subsequently this problem has
been focused further in \cite{BF 99} and \cite{LOM 98} where the
authors tried to construct the missing piece on the basis of
general arguments.

All approaches which rely on an integrated effective action
share a basic problem: The trace-anomaly is a local (UV) quantum
effect.  In order to exist as a mathematical object, the
integrated effective action assumes vanishing fields at
infinity.  But just there the flux is calculated from it. 
Therefore, already for the case of minimally coupled scalars in
general dilaton theories \cite{LVA} the CF approach was used, as
well as in the first application to nonminimally coupled scalars
by the present authors \cite{KLV 98/H}.  In that paper the same
EM conservation as in \( D=4 \) was assumed leading to an
unacceptable negative flux for SRG.

Our recent attempt to achieve an exact treatment of this problem
\cite{KV99/H} in terms of a properly adapted CF approach has
been the first successful one in the sense that it yielded the correct
relation between Hawking temperature and flux at infinity. 
Exact cancellation of the negative flux against a piece derived from a
``dilaton field anomaly'' of the effective action was observed,
leaving the  asymptotic flux for minimal coupling. 
As a by-product also the full integrated action was obtained -
but not used for this derivation, because of the problem
mentioned above.

In Section 2 we collect definitions and formulas for the exact
solution for the background, as determined by the action
(\ref{5}).  Here we also take the opportunity to clarify some
quite common misunderstandings regarding different ``conformal
frames'' for BH-s \cite{QUOTE}, although the situation has been
clear to several authors for some time \cite{KUCH}.  As a
warm-up in Section 3 we consider minimally coupled scalars in \(
D=2 \) and describe \( \zeta \)-function regularisation to
extract the trace (or ``conformal'') anomaly of the EM tensor. 
For nonminimal coupling (Section 4) the conservation of the
EM-tensor in the presence of a dilaton field receives a further
contribution \cite{BF 99},\cite{LOM 98}, which requires another
input, the variation of the effective action with respect to the
dilaton field.  The trace anomaly may be calculated for the
generalized nonminimal coupling \( \phi \rightarrow \varphi
\left( \phi \right) \) in (\ref{4}) and for a general dilaton
dependent quantum measure \cite{KLV 97/H}.  The additional
piece, the ``\( \phi \) anomaly'', no longer follows from
multiplicative variation of the differential operator in the
scalar action.  Therefore a new method had to be developed to
extract that quantity \cite{KV99/H}.  The computation of the
Hawking flux to infinity and the discussion of renormalization
which is important for the flux at finite distances in the CF
approach conclude this section.

By formal (functional) integration also the full effective
action is obtained (Section 5).  In its covariantized version it
represents the generalization of the Polyakov action for generic
dilaton theories  \cite{KV99/H}.

Section 6 is devoted to a discussion of the place where effects from
backscattering may enter the 2D approach.

We summarize our results in Section 7 where also related other
work and our own results are critically reviewed.  Open problems
and directions for further research are mentioned.

\section{Geometric background with dilaton fields}

All classical solutions of the models (\ref{5}) are 
known for some time \cite{Dil}.
They can be obtained in an especially easy way in 
the first order gravity version
of such models and using Cartan variables in the 
light-cone gauge which corresponds
to the Eddington-Finkelstein gauge for the metric 
\cite{KS92}, \cite{Kat 97}.
Letting \( \phi  \) represent one of the coordinates, 
the line element reads
\cite{Kat 97}
\begin{equation}
\label{6}
(ds)^{2}=g(\phi )\left( 2dvd\phi +l(\phi )dv^{2}\right) ,
\end{equation}
 with 
\begin{eqnarray}
 & g(\phi )=e^{-2(1-a)\phi } & \label{7} \\
b\neq -1: & l(\phi )=\frac{e^{2\phi }}{8}\left( C-\frac{2B}{b+1}e^{-2(b+1)\phi }\right) , & \label{8} \\
b=-1: & l(\phi )=\frac{e^{2\phi }}{8}\left( \tilde{C}+4B\phi \right) ,\tilde{C}=C-2B\ln 2.\label{9} 
\end{eqnarray}
 \( C \) is an integration constant, related to the ADM-mass for the subset
of such models, which are asymptotically flat.

Not all the models (\ref{5}) are suitable for analyzing Hawking
radiation.  By definition, the latter one is the amount of
energy radiated by a black hole to the asymptotic region.  That
region must represent an ``empty space'' in the solution.  There
are three subclasses of the models (\ref{6}) for which such
``empty space'' solutions are well defined.  These are the
asymptotically Minkowski (\( b=a-1 \)), asymptotically Rindler
(\( b=0 \)), and asymptotically de Sitter (\( b=1-a \)) models.

For minimally coupled scalars on such a background, all three cases
were discussed in \cite{LVA}. Here we concentrate only on the first one which
includes the \( D \)-dimensional BH-s. Bringing the solutions (\ref{7}) -
(\ref{9}) to the generalized Schwarzschild form \cite{LVA} we obtain
\begin{equation}
\label{10}
(ds)^{2}=L(u)dt^{2}-\frac{du^{2}}{L(u)}\; .
\end{equation}
 Asymptotically Minkowski models are more physically transparent, so we use
them as our main example. There we have 
\begin{equation}
\label{11}
L(u)=1-\left( \frac{u_{h}}{u}\right) ^{\frac{a}{1-a}},\qquad
\phi (u)=-\frac{1}{2(1-a)}\ln (2(1-a)u).
\end{equation}
 \( u_{h} \) is the coordinate of the horizon, defined through the equation
\( l(\phi (u_{h}))=0 \). For \( a=\frac{D-3}{D-2} \) one obtains the familiar
Schwarzschild BH in \( D \) dimensions (\( u_{h}=2M_{ADM} \) in \( D=4 \)). 

We shall also use the conformal version of 
the metric \( (x^{\pm }=\tau \pm z) \)
\begin{equation}
\label{12}
(ds)^{2}=L(u)\, dx^{+}dx^{-}=e^{2\rho \left( u\right) }dx^{+}dx^{-},\, 
\quad du=L(u)dz.
\end{equation}
It is essential, that variables are rescaled appropriately so that the function
\( L(u) \) as in (\ref{11}) tends to unity at the asymptotic region. In this
manner the introduction of a rescaling factor 
for the Hawking temperature \( T_{H} \)
is avoided. As \( L(u) \) is nothing but the Killing norm, \( T_{H} \) is
related to the surface gravity \cite{GH 76} \cite{Wald/T}
\begin{equation}
T_{H}=\frac{1}{4\pi }L^{\prime }\, \left( u\right) =\frac{1}{4\pi u_{h}}\, 
\frac{a}{1-a}   
=\frac{D-3}{4\pi u_{h}}\ , 
\label{13} 
\end{equation}
where the last expression refers to the \( D \)-dimensional SBH.

In this short review we deal only with models describing a
single Schwarzschild-like BH.  More complex solutions as, for
example, exotic multiple horizon Nariai BH-s \cite{BH/2} will not be
considered, since they are regarded as a ``pathology'' even by
the authors of corresponding research papers (see the last 
paper of ref. \cite{BH/2}). 
They also clearly are not covered by our \( 2d \) approach.

Over the last years comparisons of Hawking radiation (and of quantized
versions of \( 2d \) dilaton theories) in models differing by conformal 
transformations
\begin{eqnarray}
\label{14}
&&g_{\mu \nu }=e^{-2\xi }\, \widetilde{g}_{\mu \nu }
\\
\label{15}
&&\sqrt{-g}\, R=\sqrt{-\widetilde{g}}\, 
\widetilde{R}+2\partial _{\mu \, }\left[ \sqrt{-\widetilde{g}}\: 
\widetilde{g}^{\mu \nu }\, \partial _{\nu }\xi \right] 
\end{eqnarray}
can be found in the literature \cite{QUOTE}. Indeed, it is tempting to choose
\( \xi \left( \phi \right)  \) in (\ref{14}) in such a way that the kinetic
term \( (\nabla \phi )^{2} \) in (\ref{2}) or in the most general version
of a dilaton theory (\( X=-2\exp \) \( \left( -2\phi \right)  \))
\begin{equation}
\label{16}
\mathcal{L}_{dil}=\sqrt{-g}\left[ -\frac{XR}{2}+U(X)(\nabla
X)^{2}-V\left( X\right) \right]
\end{equation}
is made to disappear.  As a strategy to obtain a solution to
(\ref{16}) in an easier manner this is certainly legitimate -
just as the introduction of a corresponding canonical
transformation in the Hamiltonian version of such a theory.  But
it should be kept in mind that (\ref{16}), the geometric part of
the action, is \textit{not} invariant under (\ref{14}) (as it
would be for Weyl-invariance in the trivial string case).  It is
crucial, therefore, that the same transformation of fields is
performed in the ``less visible'' parts of the theory.  The
global properties of manifolds - like (\ref{11}) for the BH -
are determined by the continuation of coordinate patches through
incomplete boundaries where no singularities are located. 
``Incompleteness'' means that at least one null or non-null
geodesic reaches that boundary at a finite value of the affine
parameter.  For the BH (and for manifolds with more complicated
topology as treated by Kl\"{o}sch and Strobl \cite{KS92}) this
procedure ultimately leads to - sometimes quite complicated -
(global) Penrose diagrams.  Clearly the geodesics for the
manifold involving a certain \( g_{\mu \nu } \) are to be
computed for the \textit{same} metric.  Equivalently, those
geodesics can be considered to originate from an additional
piece of the action describing the motion of a (point) test mass
in the background geometry.  But these geodesics describing the
global properties of the manifolds must not be confused with the
couplings to matter fields as in (\ref{4}) which, of course,
remain unchanged by the formal transition from \( g \) to \(
\widetilde{g} \).

Now consider the same model after the transformation (\ref{14}),
(\ref{15}).  It is obvious from the argument above that
\textit{also in the geodesic equation} \( g \) must be
transformed to \( \widetilde{g} \)! The derivation of \( T_{H}
\) from the surface gravity, leading to (\ref{12}) had been
based upon \( g \).  Doing \textit{all} the calculations in
terms of \( \widetilde{g} \), therefore, should be consistent -
as long as no obstructions are implied by the transformation
(\ref{14}).  Because when the map \( g\rightarrow \widetilde{g}
\) is not isomorphic, or if it even introduces a singularity,
the \( \widetilde{g} \) system will become correspondingly
complicated - if treated correctly! Precisely this happens e.g.\
in the case of the DBH \cite{Eli 91} when the kinetic \( \left(
\nabla \phi \right) ^{2} \) - term of the dilaton field is
transformed away.  For that model the \( \widetilde{g} \) -
theory exhibits a geometric part with Rindler geometry (constant
acceleration) and the BH-like singularity disappears altogether
- which is clearly a completely different theory.

We stress that the experience from ordinary field theory in a
fixed (Minkowski) background is misleading.  There a
redefinition of fields does not change the (Minkowski) manifold,
upon which these fields live.  For gravity the situation is
fundamentally different.  Here the ``field'' (geometric
variable: \( g_{\mu \nu } \), dilaton field...) at the same time
defines the manifold.  Therefore even locally (cf.  the change
of the curvature, eq.\  (\ref{15})!) the transformed theory
refers to a different manifold.

As a special illustration consider the action (\ref{16}) with \( U=0 \).
Its exact solutions for the line element is \cite{KS92} (\ref{6}) with
\( g\left( \phi \right) =1 \) and, in terms of the variable \( X=u \) already
introduced in (\ref{10}), (\ref{11}),
\begin{equation}
\label{17}
l\left( u\right) =\mathcal{C}-\int _{uo}^{u}dyV\left( y\right) dy=\mathcal{C}-w\left( u\right) ,
\end{equation}
where \( C \) is the integration constant labelling a certain
solution.  It is that quantity which changes under the influence
of interacting matter \cite{KS92}.  Such a model can only
describe ``eternal'' singularities, because they must appear in
\( w(u) \) and are determined by the (given) parameters of the
action.  Some instantaneous influx of matter only shifts the
position of the horizon and has no effect upon the singularity
(-ies).  Outside the horizon we may identify \( u \) with the
``radius''.  The surface gravity allows an interpretation as a
Hawking temperature at a horizon, if \( w\left( \infty \right)
\rightarrow 0 \), i.e.  if the metric (\ref{6}) becomes
asymptotically flat \footnote{
asymptotics may be discussed in the similar manner.  }.  After a
conformal transformation (\ref{14}) with \( w ( u ) = \exp 2\xi
\) and after introducing a new coordinate \( d \tilde{u}=du\,
w^{-1}\left( u\right) \) the line element \( \left( ds\right)
^{2} \)with (\ref{17}) and the new one
\begin{eqnarray}
\label{18} &&
\left( d\widetilde{s}\right) ^{2}=2\,
d\widetilde{u}dv+\widetilde{l}\left( \widetilde{u}\right)
(dv)^{2}
\\
\label{19} &&
\widetilde{l}\left( \widetilde{u}\right)
=\frac{\mathcal{C}}{w(u(\widetilde{u}))}-1
\end{eqnarray}
can be compared.  Clearly (\ref{18}) with (\ref{19}) will be
solutions of a model (\ref{16}) with \( U\neq 0 \).  But in
terms of the new radial variable \( \widetilde{u} \) now
obviously a new singularity develops at \(
\widetilde{u}\rightarrow \infty \) (or \( u\rightarrow -\infty ,
\) depending on the sign of \( w \)).  On the other hand, the
original ``eternal'' singularity in \( w(u) \) generically will
disappear, but new ones will develop from the zeros of the
latter quantity.  The original horizon at \( u=u_{h} \) in
(\ref{17}) still will be a horizon (zero) of (\ref{19}), but it
will be related to quite different singularities, and - if such
a region exists at all for the new model (\ref{18})! - a
different asymptotically flat regime at the, in general,
opposite side of the horizon.  A theory with generic metric
(\ref{19}) clearly has a flat ground-state (\( \mathcal{C}=0
\)).  SBH and DBH are special cases.  In fact, the Lagrangian
(\ref{16}) for a general dilaton theory with this ``Minkowski
ground state'' property, where \( U \) depends in a specific way
on \( V \), can be written down easily \cite{Kat 97}.

Additional complications by the field transformation (\ref{14}),
(\ref{15}) arise in the quantum version of such theories .  E..g. 
due to the change of global properties the definition of
``asymptotic states'' for some quantum gravity S-matrix ( a
horrendously difficult problem by itself) becomes completely
undefined.  Thus ``quantization'' in terms of \( g \) resp.\ \(
\widetilde{g} \) (or using even locally well-defined canonical
transformations in a Hamiltonian formulation) refers to
completely different quantum field theories \cite{KUCH}.

In view of these comments it cannot come as a surprise that the
ADM mass of a black hole is not conformally invariant
\cite{Chan,LVA}.  As in the case of 4D Einstein gravity, in
order to define the ADM mass unambiguously it is crucial to
choose a proper asymptotic behavior of the metric \cite{FF}. 
Again, it is important which metric, $g$ or $\tilde g$, is the
``physical'' one.  The very existence of the ADM mass means that
the action is not invariant under the diffeomorphisms which do
not vanish at the boundary.  Another example of conformal
non-invariance of physical processes in $2D$ dilaton gravity has
been given recently in \cite{cruzfabbri}.

\section{Hawking radiation from minimally coupled scalars}

\subsection{EM tensor}

The EM tensor for minimally coupled scalars is defined from the
action \( W^{(min)} \) with the Langrangian (\ref{3}), dropping
the dilaton factor \( e^{-2\phi } \) :
\begin{equation}
\label{20}
T^{\left( min\right) }_{\mu \nu
}=\frac{2}{\sqrt{-g}}\frac{\delta W^{\left( min\right) }}{\delta
g^{\mu \nu }}= \left( \partial _{\mu }f\right) \left( \partial
_{\nu }f\right) -\frac{g_{\mu \nu }}{2}\left( \partial ^{\alpha
}f\right) \left( \partial _{\alpha }\, f\right)
\end{equation}

{}From diffeomorphism invariance of the matter action (see below
for the more general argument, if dilaton fields are present in
the Langrangian (\ref{4})), taking the matter fields on-shell,
the usual EM conservation follows (we drop the superscript (min)
for simplicity in this section):
\begin{equation}
\label{21}
\nabla _{\mu }T^{\mu \nu }=0 \ .
\end{equation}

In the conformal gauge (\ref{12}) only the components \( \Gamma
_{++}\, ^{+}=2\partial _{+}\rho ,\, \Gamma _{--}\,
^{-}=2\partial _{-}\rho \) are nonzero.Thus equation (\ref{21})
for \( \nu =+ \) reads 
\begin{equation}
\label{22}
\partial _{+}T_{--}+\partial _{-}T_{+-}-2(\partial _{-}\rho 
)T_{+-}=0\; .
\end{equation}
 According to (\ref{11}), (\ref{12}) the background depends on the variable
\( u \) alone. Thus derivatives of light cone coordinates, acting on functions
of \( u \) become 
\begin{equation}
\label{23}
\partial _{+}=-\partial _{-}=\frac{1}{2}\partial _{z}=-\frac{1}{2}L(u)
\partial _{u}\, ,
\end{equation}
and (\ref{22}) turns into a simple first order differential equation
in \( z \) 
\begin{equation}
\label{24}
\partial _{z}T_{--}=\left[ \partial _{z}-2\left( \partial
_{z}\rho \right) \right] T_{+-},
\end{equation}
which may be integrated easily for the flux component \( T_{--}
\)~if \( T_{+-} \)~~is known.  We will be interested in the flux
at infinity where the space should become flat (\( g_{\mu \nu
}\rightarrow \eta _{\mu \nu } \)~).  In light-cone coordinates
\( x^{\pm }=x^{\circ }\pm x^{1}=\tau \pm z \) the flux
orthogonal to the lines \( x^{-}=const.  \) in the direction of
outgoing waves is just \( T_{--} \).  On the other hand, \(
T_{+-\, } \) is nothing else but the trace of \( T_{\mu \nu } \) ,
\begin{equation}
\label{25}
T^{\mu }\, _{\mu }=g^{\mu \nu }\, T_{\mu \nu } = 
4e^{-2\rho } T_{+-} , 
\end{equation}
which vanishes identically for the classical action according to
(\ref{20}).  A solution for \( f \) in conformal gauge still
obeys the free wave equation \( \partial _{+}\partial _{-}f\, =0
\) with solutions \( f=F_+\left( x^{+}\right) +F_-\left(
x^{-}\right) \).  Thus the EM term (\ref{20}) at the classical
level at large distances \( f=F_\pm \) is of the form
\begin{equation}
\label{26}
\left. T_{\mu \nu }^\pm \right|_{as} = a_\pm
 \left( _{\pm 1\, \, \, \, \, 1}^{1\, \, \, \, \, \pm 1}\right) , 
\end{equation}
where \( a_{\pm }\propto \left( F_{\pm }^{\prime }\right)^{2}
\).  Classically no flux to infinity should occur from a BH,
therefore \( a_\pm = 0 \) which is consistent with the trivial
solution ~\( f\, =0 \).

On the other hand, if Hawking radiation exists, it must be a
quantum effect.  Then the constant \( a_\pm =\frac{\pi }{12}\,
T^{\, 2}_{H} \) (the Stefan-Boltzmann law in \( D=2 \)) in front
of (\ref{26}) can be determined by comparison with the EM tensor
\( T^{\, \left( e\right) }_{\mu \nu } \)for black body radiation
in \( D=2 \) in equilibrium with a heat bath, where the pressure
equals the radiation density,
\begin{equation}
\label{27}
T^{\, \left( e\right) }_{\mu \nu }\mid _{as}=\frac{\pi }{6}T^{\,
2}_{H}\, \left( _{0\, \, \, 1}^{1\, \, \, 0}\right) .
\end{equation}

The equilibrium EM tensor from (\ref{26}) with $a_+ = a_-$ must
be \( T^{+} + T^{-} \) \cite{Chris 77}.  In any case, for the
determination of the flux at infinity it is enough to know one
component, like \( T_{--} \)~introduced above.  As emphasized
already, minimally coupled scalars in \( D=2 \)~conformal gauge
(\ref{12}) also at nonasymptotic distances are still determined
by a free wave equation.  They do not ``feel'' the geometry. 
Therefore \( T_{\mu \nu }\mid _{as} \) will coincide with the
asymptotic value of \( T_{\mu \nu } \)~as calculated from the
classical action.  This is not the case in \( D=4 \) where a
complication from emission - absorption coefficients
(``grey-factors'') arises \cite{Chris 77}, \cite{Muk 94}, and,  as
will be sketched below (Section 6), also introduces
complications in the spherically reduced case, i.e.  for
nonminimally (dilaton coupled) scalars.

If first order quantum loops of the scalar field are taken into
account, the full (effective) action \( W \) will acquire a
contribution with nonvanishing trace (\ref{25}).  As it breaks
the conformal symmetry of the scalar interaction - note, however
that the geometric part of the action is not conformally
invariant! -, this new term will be called the ``conformal
anomaly''.

We now interpret (\ref{24}) to represent the one-loop quantum
contributions on both sides of this equation.  If the conformal
anomaly is known, it may be integrated outside the horizon \(
u_{h}\leq u\, \left( -\infty \leq z\right) \, \)all the way to
its asymptotic value \( T_{--}\mid _{u\rightarrow \infty
}=T_{--}\mid _{as} \) .  Beside \( T_{+-} \) the only input then
is an integration constant which determines \( T_{--}\mid
_{u_{h}}=t_{--} \) .

Before discussing different choices of the boundary values for
\( T_{--}\mid _{u_{h}} \)which are related to different
``quantum vacua'' \cite{BOU 76}, \cite{ISR 76},\cite{UN 76} it
is important to understand the behaviour of \( T_{--}\mid
_{u_{h}} \)in terms of global (Kruskal - Szekeres) coordinates. 
For a nondegenerate Killing horizon in the neighbourhood of \(
u\approx u_{h} \)~ the Killing norm has a simple zero.  Hence in
(\ref{5})
\begin{equation}
\label{28}
\left( ds\right) ^{2}\approx 2dv\left( du+a\left( u-u_{h}\right) dv\right) 
\end{equation}
with some constant \( a \), proportional to the surface gravity
(which in the following will be set equal to \( 1 \)),  in the
Killing norm \( l\, \left( u\right) \approx a\left(
u-u_{h}\right) \).  The variable \( u \) is taken to be the one
which is related to the conformal radial variable \( z \)
(tortoise coordinate) in (\ref{11}),
\begin{equation}
\label{29}
z=\int ^{u}\frac{dy}{l(y)}\approx \ln  
\left( u-u_{h}\right) \approx \ln l\left( u\right) ,
\end{equation}
so that in conformal gauge (\ref{12}) \( \left( x^{-}=\tau -z\right)  \)
\begin{equation}
\label{30}
\left( ds\right) ^{2}\approx e^{z}dx^{+}dx^{-}=e^{\left( x^{+}-x^{-}\right) }dx^{+}dx^{-}.
\end{equation}
 Global conformal coordinates \( \left( ds^{2}\approx d\bar{x}^{+}d\bar{x}^{-}\right)  \)~near
the horizon are introduced by
\begin{equation}
\label{31}
\bar{x}^{\pm }=\pm 2\, e^{\pm x^{\pm }/2}\, ,
\end{equation}

so that the components of \( \bar{T}_{\mu \nu } \) by
reparametrization \( x^{\pm }\rightarrow \bar{x}^{\pm }
\)become
\begin{eqnarray}
\bar{T}_{--} & = & \frac{4}{\left( y^{-}\right) ^{2}}\, T_{--},\nonumber \\
\bar{T}_{+-} & = & -\frac{4}{y^{+}y^{-}}\, T_{+-},\label{32} \\
\bar{T}_{++} & = & \frac{4}{\left( y^{+}\right) ^{2}}\, T_{++}.\nonumber 
\end{eqnarray}
For the factor of \( T_{--} \)  in \( \bar{T} \)\( _{--} \)
at fixed \( x^{+}=\tau +z \) we obtain
\begin{equation}
\label{33}
\left( y^{-}\right) ^{-2}=\frac{1}{4}e^{\, \left( \tau -z\right) }
=\frac{1}{4}e\, ^{\left( x^+ -2z\right) }\propto \left( u-u_{h}\right) ^{-2}.
\end{equation}
In a similar way the factors in \( \bar{T}_{-+} \)  and \(
\bar{T}_{++} \) are found to be \( \left( u-u_{h}\right)
\)\( ^{-1} \) and a constant, respectively.

Before it had been realized that BH-s may radiate, it had seemed
natural enough to assume that \( T_{--}\mid _{as}\rightarrow 0
\) ( Boulware vacuum \cite{BOU 76}).  This leads to a divergent
\( \bar{T}_{--} \) at the horizon \cite{Chris 77}.  Another
proposal considering the BH-system in thermal equilibrium with a
heat bath \cite{ISR 76,GH 76} assumed regularity of \( T_{\mu
\nu } \) at the past \textit{and} at the future event horizon. 
In this case (\ref{27}), the sum of ingoing and outgoing fluxes
at infinity would be relevant.  This is not the situation we are
interested in.  However, from the argument above which is
restricted to the region between the future horizon and positive
null infinity \( \mathcal{I}^{+} \)(\( u=\infty ) \) the choice
\( T_{--}\mid _{u\approx u_{h}}=O \) \( \left( u-u_{h}\right)
^{2} \), providing a finite flux in global coordinates at the
horizon \cite{UN 76}, seems the most obvious one.

Indeed in ref.\  \cite{Chris 77} it was shown that the Unruh
vacuum for \( D=2 \) minimally coupled scalars is consistent. 
It precisely relates the Hawking temperature \( T_{H}=\left(
4\pi u_{h}\right) ^{-1} \)to the asymptotic flux (\ref{21}) when
the (known) conformal anomaly \( T_{+-\, } \)(see below) is
used.

\subsection{Conformal anomaly}

The contribution of the scalar loop to the trace of the EM 
tensor in an external
gravitation field is most effectively calculated in the heat-kernel approach
\cite{Atiyah},\cite{Gilkey}.

The classical action from (\ref{5}) for minimal coupling by 
partial integration can be written as 
\begin{equation}
\label{34a}
S^{(min)} = \frac{1}{2}\; \int\, \sqrt[4]{-g}\, f\, A^{(min)} 
\, \sqrt[4]{-g}\, f
\end{equation}
with the differential operator 
\begin{equation}
\label{35a}
A^{(min)} \; = \; - g^{\mu\nu}\, \nabla_\mu\, \nabla_\nu\; ,
\end{equation}
pulling determinants of the metric through the covariant 
derivative $\nabla_\mu$ . In the path integral for the one-loop 
effective action $W$ we turn to the Euclidean region $\sqrt{-g} 
\to \sqrt{g}$ \cite{textbook}
\begin{equation}
\label{36a}
\exp W = \int\, (df\, \sqrt[4]{g})\, \exp\, S\, ,
\end{equation}
where the factor $ \sqrt[4]{g}$ in the measure guarantees 
covariance of the Gaussian measure 
\begin{eqnarray}
\label{37a} &&
\int\, (\sqrt[4]{g}\, df)\; \exp\; \langle\, f, \, f\, 
\rangle \; = \; \mbox{invariant}
\\
\label{38a} &&
\langle \, f_1, \, f_2\, \rangle \; = \; 
\int\, d^2\, x\, \sqrt{g}\, f_1\, f_2
\end{eqnarray}
The result of the integral (\ref{36a}) is the effective action, 
 given by the functional determinant of \( A^{(\rm min)} = A \) 
\cite{textbook} 
\begin{equation}
\label{34}
W = \ln\, (\det\, A)^{1/2} = \frac{1}{2}\,\mbox{Tr}\,\ln A\quad .
\end{equation}
In the zeta-function regularization \cite{zeta} a power of \( A \) instead
of the logarithm is considered: 
\begin{equation}
\label{35}
W = -\frac{1}{2}\zeta'_{A}\,(0),\qquad \zeta _{A}(s)={\textrm{Tr}}(A^{-s})
\end{equation}
Prime denotes differentiation with respect to \( s \). No analytic general
results exist for the finite part of (\ref{35}), but they do for certain 
variations
of this quantity. We start our analysis with the calculation for an \textit{arbitrary}
conformally covariant operator \( A \).

Conformal covariance means that under an infinitesimal conformal transformation
of the metric \( \delta g_{\mu \nu }=\delta k(x)g_{\mu \nu ,}\delta g^{\mu \nu }=-\delta kg^{\mu \nu } \)
the operator \( A \) transforms as 
\noindent 
\begin{equation}
\label{36}
\delta A=-\delta kA.
\end{equation}
The same conformal transformation in the effective action produces
the quantum contribution to 
trace of the energy-momentum tensor~~(\ref{20}) 
\begin{equation}
\label{37}
\delta W=\frac{1}{2}\int d^{2}x\sqrt{g}\delta g^{\mu \nu }T_{\mu \nu }=-\frac{1}{2}\int d^{2}x\sqrt{g}\delta k(x)T_{\mu }^{\mu }(x).
\end{equation}
The variation of the zeta function with respect to a certain
parameter or field in \( A \) is related to the one of the
operator \( A \) \cite{EKP} \cite{Gilkey} : 

\begin{equation}
\label{38}
\delta \zeta _{A}(s)=-s{\textrm{Tr}}((\delta A)A^{-1-s})
\end{equation}

Due to conformal covariance (\ref{36}) the powers of \( A \) in
(\ref{38}) recombine back into \( A^{-s} \).  Thus with the
definition of a generalized \( \zeta \)-function \cite{Atiyah}
\cite{Gilkey}
\begin{equation}
\label{39}
\zeta \left( s\vert \delta k, A) = {\rm Tr}\, (\delta kA^{-s}\right) 
\end{equation}
the variation in (\ref{37}) can be identified with
\begin{equation}
\label{40}
\delta W=-\frac{1}{2}\zeta (0|\delta k,A).
\end{equation}
We thus encounter a \textit{multiplicative} variation of the operator
\( A \). For such variations of \( \zeta _{A} \) the general result is known.

Combining (\ref{40}) and (\ref{37}) yields
\begin{equation}
\label{41}
\zeta (0|\delta k,A)=\int d^{2}x\sqrt{g}\delta k(x)T_{\mu }^{\mu }(x).
\end{equation}
By a Mellin transformation one can show that \( \zeta (0|\delta k,A)=a_{1}(\delta k,A) \)
\cite{Gilkey}, where \( a_{1} \) is defined as a coefficient in a small \( t \)
asymptotic expansion of the heat kernel containing \( F \) multiplicatively:
\begin{equation}
\label{42}
{\textrm{Tr}}(F\exp (-At))=\sum _{n}a_{n}(F,A)t^{n-1}
\end{equation}
To evaluate the form of \( a_{1} \)  we use the standard
method \cite{Gilkey}. We assume that \( A \) as in our 
application 
is an operator of Laplace type.
This means that one can represent it as 
\begin{equation}
\label{43}
A=-\left( \hat{g}^{\mu \nu }\hat{D}_{\mu }\hat{D}_{\nu }+E \right) ,
\end{equation}
 with a suitable choice of the metric \( \hat{g}^{\mu \nu } \), a related covariant
derivative \( \hat{D}_{\mu } \), and an endomorphism \( E \). Then the result
for \( a_{1} \) can be simply taken from the literature \cite{Gilkey},
returning to Minkowski space ($\sqrt{g} \to \sqrt{-g}$)
\begin{equation}
\label{44}
a_{1}(\delta k,A)=\frac{1}{24\pi }{\textrm{tr}}\,\int d^{2}x\sqrt{-\hat{g}}
\,\delta k(\hat{R}+6E)\; .
\end{equation}
 Here \( \textrm{tr} \) denotes ordinary trace over all matrix indices (if
there are any in \( A \)). \( \hat{R} \) is the scalar curvature for the metric
\( \hat{g} \). 

For \( A = A^{(min)} \) in (\ref{35}), 
 \( g^{\mu \nu } = \hat{g}^{\mu \nu },\,
E=0 \) the conformal anomaly follows immediately from (\ref{44})
and (\ref{37}) with the definitions (\ref{20}) for the EM tensor
now used for the \textit{effective} action (\ref{34})

\begin{equation}
\label{45}
T_{\mu }^{\mu }=\frac{1}{\sqrt{g}}\frac{\delta }{\delta
k}a_{1}(\delta k,A^{(min)})=\frac{1}{24\pi }R.
\end{equation}
The Ricci scalar \( R \) is most easily computed for the
conformal gauge by means of the identity (\ref{15}) (\( \xi
=-\rho , \) \( \widetilde{g}_{\mu \nu }=\eta _{\mu \nu
,}\widetilde{R}=0) \)
\begin{equation}
\label{46}
R=-\partial _{+}\partial _{-}\rho ,
\end{equation}
 yielding the \( (+-) \) component of the energy-momentum tensor 
 (cf.\ (\ref{25}))
\begin{equation}
\label{47}
T^{(min)}_{+-}=-\frac{1}{12\pi }\partial _{+}\partial _{-}\rho
=\frac{1}{48\pi }\partial _{z}^{2}\rho
\end{equation}

\subsection{Radiative flux}

Solving eq.\ (\ref{27}) the radiative flux is obtained
with the anomaly (\ref{47})
\begin{equation}
\label{48}
T_{--}^{\left( min\right) }
= \frac{1}{48\pi }\left[ \partial_z ^{2}\rho -\left( \partial
_{z}\rho \right) ^{2}\right] +t_{--}\; ,
\end{equation}
where  \( t_{--} \) is the integration constant.  

In terms of (\ref{12}) this may be rewritten as
\begin{equation}
\label{49}
T_{--}^{\left( min\right) }=\frac{1}{48\pi }\left[
\frac{L^{\prime \prime }L}{2}-\frac{{L^\prime}^{2}}{4}\right]
+t_{--} \; .
\end{equation}
The behaviour of \( L( u) \) near \( u\approx u_{h} \) can be
read off from the exact result for a BH background 
\begin{equation}
\label{50}
T_{--}^{\left[ min\right] }=t_{--}-\frac{1}{192\pi u^{2}_{h}}
+\frac{\left( u-u_{h}\right) ^{2}}{192\pi u^{4}u_{h}^{2}}\left(
u^{2}+2uu_{h}+3u_{h}^{2}\right) \ .
\end{equation}
One observes that a finite flux \( \bar{T}_{--} \)in global
coordinates (\ref{32}) at the horizon 
is indeed provided by fixing \( t_{--} \) \textit{alone}
to cancel the first term. On the other hand, the asymptotic flux at \( u\rightarrow \infty  \)
is just given by the same expression as \( t_{--}\,  \)when the other terms
in (\ref{49}) tend to zero,  
\begin{equation}
\label{51}
T^{(min)}_{--}|_{asymp}=t_{--}=\frac{\pi }{12}T_{H}^{2},
\end{equation}
a result which is in perfect agreement with (\ref{26}).

One can show \cite{LVA} that the expression (\ref{51}) for \(
T_{--} \) in terms of the Hawking temperature is valid also for
asymptotically Rindler and de Sitter models.  In this connection
we emphasize that the calculation of the Hawking flux is
sensitive to the choice of the coordinate system.  It is
essential that \( \rho \rightarrow 0 \) in the asymptotic
region.  For \( \rho \rightarrow const\neq 0 \) the flux is
measured using time and length scales different from the ones
used to measure the BH mass.  If \( \rho \) does not tend to a
constant in the asymptotic region, the observer connected with
such a coordinate system measures a mixture of Hawking and Unruh
radiation.  This happened in a recent calculation \cite{Kim} for
AdS BH-s, where the puzzling result of zero Hawking flux was
obtained.  Note, that due to the condition \( \phi =0 \) used in
that reference to obtain solutions of the quantum corrected
field equations the properly normalized (classical) ADM mass
(cf.  eq (\ref{14}) in \cite{LVA}) would be zero.

So far in this section only models with one horizon were
discussed.  But the only necessary input for the method
employed, the integration of the EM conservation in the sense of
ref.\  \cite{Chris 77} between the future event horizon and
(flat) null-infinity \( \mathcal{I}^{+} \) applies equally well
for all dilaton theories (\ref{16}) provided their Killing norm
\( L\left( u\right) \) in the interval \( u_{h}\leq u \)\( \leq
\infty \: \)also behaves like (\ref{11}), i.e.\  \( L(\infty )=L(u_{h})=0
\) when \( u_{h} \) is the largest (simple) zero.  That horizon
may protect more than one singularity - or even none at all.  We
just note that within the first order gravity formulation of \(
2D \) theories \cite{KS92} it is straightforward to ``design''
such models easily by appropriate choices of \( U(X) \) and \(
V(X) \) in (\ref{16}), where \( U(X) \) corresponds to a
``torsion term'' in the equivalent Cartan theory.

\section{Nonminimally coupled scalars}

As explained in the Introduction this case should be the closest
one to the situation in (higher dimensional) Einstein gravity. 
Indeed the spherically reduced action (\ref{2}) together with
(\ref{4}) exactly reproduces the same equations of motion, which
are obtained from the variation of the Einstein-Hilbert action
in \( D \) dimensions when the dependence of the metric on \(
x^{\mu }=\left( t,r\right) \) alone, as in (\ref{1}), is assumed
\cite{Gru 99}.  This is not a trivial result: e.g.  reducing
with an ansatz of a ``warped'' metric does not allow the
definition of a reduced Lagrangian from which the entire set of
e.o.m.-s follow \cite{WARP}.

\subsection{EM tensor}

With the same definition (\ref{20}) of the EM tensor \( T_{\mu \nu }^{(nm)}
\) for the nonminimal case, in the presence of
an additional dilaton field we obtain at the classical level \(
T_{\mu \nu }^{\left( nm\right) }=e^{-2\phi }T^{(min)}_{\mu \nu }
\).  The factor \( e^{-2\phi } \) is nothing else but the (for
\( D\neq 4 \) appropriately redefined) scaling factor or
(radius)\( ^{2} \) (cf.\ (\ref{1}) for SRG).  If interpreted
with respect to the unreduced dimensional level, the flux is
multiplied by the area of the sphere \( S^{D-2} \)(remember that
\( \int d^2\Omega \) has been integrated out) - as it should be. 
In the present case where \( W^{\left( nm\right) }=W^{\left(
nm\right) }\left( g_{\mu \nu },f,\phi \right) \) the
conservation law (\ref{24}) must be modified.  By construction
the matter field action is invariant under the diffeomorphism
transformations
\begin{eqnarray}
\delta g_{\mu \nu } & = & \nabla _{\mu }\xi _{\nu }+\nabla _{\nu
}\xi _{\mu ,}\nonumber \\ 
\delta \phi & = & \xi ^{\nu }\partial
_{\nu }\phi ,\label{52} \\ 
\delta f & = & \xi ^{\nu }\partial_{\nu } f\; ,\nonumber
\end{eqnarray}
where \( \phi \) denotes either the dilaton field or any local
function thereof.  By applying the transformations (\ref{52}) to
the nonminimal action \( W \)\( ^{(mn)} \) one obtains instead
of (\ref{21})
\begin{equation}
\label{53}
\nabla ^{\mu }T^{\left( nm\right) }_{\mu \nu }= -(\partial _{\nu
}\phi )\frac{1}{\sqrt{-g}}\frac{\delta W^{(nm)}}{\delta \phi },
\end{equation}
where a term containing \( \delta W^{(mn)}/\delta f \) has been
dropped on the r.h.s., when we assume the fields \( f \) to
be on-shell classically.  Having integrated out the fluctuation
of these fields in the one loop order also no contribution can
come from those when (\ref{53}) is applied to the effective
action.  When Hawking radiation is to be studied from (\ref{53})
we shall assume that we start from a situation where no
classical field \( f \) is present, although some modifications
will turn out to be necessary in a more detailed consideration
(see Section 6).  Then the r.h.s.\  of (\ref{53}) vanishes for
its classical contribution.  As \( T^{\left( nm\right) } \) also
has a classically vanishing trace the situation is much like the
minimal case (53), except for the fact that beside the conformal
anomaly \( T_{+-}^{\left( nm\right) } \) now also another term
(``\( \phi \)-anomaly'') appears together with contributions from 
the 1-loop version of (\ref{53}), viz.
\begin{equation}
\partial _{+}T_{--}^{\left( nm\right) }  
=  -\left( \partial_{-}-2\left( \partial _{-}\rho \right) \right) 
T_{+-}^{\left( nm\right) }\, 
 +\frac{( \partial _{-}\phi )}{\sqrt{-\eta }}
\frac{\delta W^{\left( nm\right) }}{\delta \phi }
\label{54} 
\end{equation}
or (cf. (\ref{23}))
\begin{equation}
\partial _{z}T_{--}^{\left( nm\right) } = \left( \partial
_{z}-2\left( \partial _{z}\rho \right) \right) T_{+-}^{\left(
nm\right) } -\frac{\partial _{z}\phi }{\sqrt{-\eta}}
\frac{\delta W^{\left( nm\right) }}{\delta \phi }.\label{55}
\end{equation}
The last term in  (\ref{55}) had not been
taken into account in \cite{KLV 97/H}.

Separating in \( T_{+-}^{\left( nm\right) } \) the contribution of
minimal coupling from the additional \( T^{\left( dil\right) }_{+-} \) which
appears for non-vanishing dilaton fields
\begin{equation}
\label{56}
T_{+-}^{\left( nm\right) }=T_{+-}^{\left( min\right)
}+T_{+-}^{\left( dil\right) } \; ,
\end{equation}
in the total \( T_{--}^{\left( nm\right) } \) from integration of (\ref{55})
we obtain three pieces where the first one is known from (53):
\begin{eqnarray}
\label{57} &&
T_{--}^{\left( nm\right) }=T_{--}^{\left( min\right) }+T_{--}^{\left( dil\right) }+T_{--}^{\left( \phi \right) }
\\
\label{58} &&
T_{--}^{\left( dil\right) }=\int _{-\infty }^{z}\left[ \left( \partial _{z}-2\left( \partial _{z}\rho \right) \right) T^{\left( dil\right) }_{+-}\right] _{z^{\prime }}\, dz^{\prime }
\\
\label{59} &&
T_{--}^{\left( \phi \right) }=
\int _{-\infty }^{z}dz^{\prime }\left( \partial _{z^{\prime }}\phi \right) 
\left( \frac{1}{\sqrt{-\eta}}
\frac{\delta W^{\left( nm\right) }}{\delta \phi }\right) _{z^{\prime }}.
\end{eqnarray}
All integrals start at the horizon (\( z=-\infty  \) or \( u=u_{h} \)).
By analogy with the minimal case \( T_{--}\vert_{u_{h}} = 0 \) is assumed and,
therefore, the overall integration constant \( t_{--} \) has been dropped.

\subsection{Conformal anomaly}

We again follow the steps outlined in section 3.3, but starting from the nonminimal
Lagrangian (\ref{4}). This will change the differential operator \( A \) to
be used in (\ref{34}). 

A less trivial observation is that the dilaton field \( \phi \)
enters also the path integral measure (\ref{31}).  This happens
because the scalar product for the \( s \)-wave modes in \( D \)
dimensions contains \( \phi \), so that instead of (\ref{38a}) 
\begin{equation}
\label{60}
<f_{1},f_{2}>=\int d^{2}x\sqrt{-g}e^{-2\phi }f_{1}f_{2}
\end{equation}
holds. Therefore, only the redefined fields 
fields \( \tilde{f}=e^{-\phi }f \) in two dimensions have a 
scalar product like (\ref{38a}). Consequently, $\tilde f$ 
possesses the  Gaussian path integral measure (\ref{37a}).  In the
following we allow for a more general coupling and a more
general path integral measure.  To this end we replace \( \phi
\) by $ \varphi ( \phi ) $ in the action (\ref{5}) and \( \phi
\) by $ \psi ( \phi ) $ in the measure, assuming that \(
\varphi \) and \( \psi \) are some local functions of the
dilaton.

Expressing the classical action corresponding to the Lagrangian
(\ref{5}) with \( \phi \rightarrow \varphi \left( \phi \right) \)
in terms of the field \( \tilde{f} \) yields the action

\begin{equation}
\label{61}
S^{(nm)} = \frac{1}{2}\int \sqrt{-g}d^{2}x\, \,
\tilde{f}A^{(nm)}\tilde{f},
\end{equation}
 containing the operator 
\begin{eqnarray}
A^{(nm)} & = & -e^{-2\varphi +2\psi }g^{\mu \nu }\left( \nabla
_{\mu }\nabla _{\nu }+2(\psi _{,\mu }-\varphi _{,\mu })\nabla
_{\nu }+\psi _{,\mu \nu }-\right.  \nonumber \\ & & \left.  -
2\varphi _{,\mu }\psi _{,\nu }+ \psi _{,\mu }\psi _{,\nu
}\right)\; .
\label{62}
\end{eqnarray}
Thus the basic quantities entering the standard form (\ref{43}) of the Laplace
type operator are 
\begin{equation}
\label{63}
\hat{g}^{\mu \nu }=e^{-2\varphi +2\psi }g^{\mu \nu }\, \, ,\qquad E=\hat{g}^{\mu \nu }(-\varphi _{,\mu }\varphi _{,\nu }+\varphi _{,\mu \nu }),
\end{equation}
 where \( D_{\mu }=\nabla _{\mu }+\omega _{\mu } \), \( \omega _{\mu }=\psi _{,\mu }-\varphi _{,\mu } \).
{}From equation (\ref{44}) we can immediately read off the conformal anomaly
for this case: 
\begin{equation}
\label{64}
T_{\mu }^{\mu }=\frac{1}{24\pi }(R-6(\nabla \varphi )^{2}+4\Box
\varphi +2\Box \psi )\, .
\end{equation}
The expression (\ref{64}) for the conformal anomaly in a dilaton theory with
general nonminimal coupling 
\( \exp\left( -2\varphi  \left( \phi \right) \right)  \)
to the scalars and for a general dilaton dependent norm involving 
\( \exp\left( -2\psi \left( \phi \right) \right)  \)
was given first by the authors together with H. Liebl \cite{KLV 97/H}. For
the special case SRG (\( \varphi =\psi =\phi  \)) it appeared already in \cite{Muk 94}.
However, another recent computation in the latter case \cite{BH 97/1} missed
two important ingredients and therefore arrived at an incorrect result: the
modification of the norm (\ref{60}) and the omission of a null-mode for the
compact manifold \cite{DOW 98} upon which the computation of the anomaly had
been performed. In the language of our much simpler approach, using well-known
heat-kernel techniques (\cite{zeta}, \cite{Atiyah}, \cite{EKP}) with local
scaling function the corresponding total derivatives cannot be missed. Although
the situation was clarified soon \cite{KLV 98/H} the incorrect factor in front
of \( \Box \varphi =\Box \psi =\Box \phi  \) from the 
last two  terms in (\ref{64})\cite{BH 97/1}
has been quoted and /or used in several papers \cite{BH/2}. It is by no means
``ambiguous'' \cite{BH 97/1}, \cite{Mik 98}.

The first term on the right hand side had been encountered already
in the case of minimal coupling. Therefore, in the notation (\ref{56})
in the conformal gauge
\begin{equation}
\label{65}
T_{-+}^{(dil)}=\frac{1}{12\pi }\left[ 2\partial _{+}\partial
_{-}\varphi -3\left( \partial _{+}\varphi \right) \left(
\partial _{-}\varphi \right) +\partial _{+}\partial _{-}\psi
\right] ,
\end{equation}
and the contribution (\ref{58}) to \( T_{--}^{\left( nm\right) } \)
becomes 
\begin{equation}
\label{66}
T_{--}^{(dil)}=-\frac{1}{48\pi }\int _{-\infty }^{z}dz^{\prime
}\left( \partial _{z^{\prime }}-2\partial _{z^{\prime }}\rho
\right) \left[ 2\partial _{z}^{2}\varphi -3\left( \partial
_{z}\varphi \right) ^{2}+\partial _{z}^{2}\psi \right]
_{z^{\prime }} \ .  
\end{equation}

\subsection{\noindent \protect\( \Phi \protect \)-anomaly}

The variation of the effective action with respect to \( \varphi  \) or \( \psi  \)
does not exhibit the same multiplicative property as the conformal variation,
because after substituting in (\ref{38}) the variation of \( 
A^{(nm)} \)  of (67) does not
recombine to powers of \( A^{(nm)} \).  Therefore, the heat kernel technique is not
applicable to the evaluation of (67) as it stands. However, crucial simplifications
occur after relating it to flat space by means of a functional integral in conformal
gauge \( g_{\mu \nu }=e^{2\rho }\eta _{\mu \nu } \). There one has \( W^{\left( nm\right) }=W^{\left( nm\right) }(\rho ,\phi ) \)
and thus the identities  
\begin{eqnarray}
\label{67} 
\frac{\delta W^{\left( nm\right) }(\rho ,\phi )}{\delta \phi
} \; & = & \; \int _{0}^{\rho }d\sigma \frac{\delta ^{2}W^{\left( nm\right)
}(\sigma ,\phi )}{\delta \sigma \, \delta \phi }+\frac{\delta
W^{\left( nm\right) }(0,\phi )}{\delta \phi }\; , \\
\label{68} 
\frac{\delta W^{\left( nm\right) }\left( 0,\phi \right) }{\delta
\phi } \; & = & \; \frac{\delta W^{(nm)}\left( 0,\varphi ,\psi \right)
}{\delta \varphi }\frac{d\varphi }{d\phi }+\frac{\delta
W^{\left( nm\right) }\left( 0,\varphi ,\psi \right) }{\delta
\psi }\frac{d \psi }{d \phi }
\end{eqnarray}
are obvious.  They relate the variation to the one at \( \rho =0
\), i.e.  to a flat background.  On the other hand, the first
term on the right hand side of (\ref{67}) can be expressed in
terms of the trace of the EM tensor (conformal anomaly) by
(\ref{37}):
\begin{equation}
\label{69}
\frac{\delta W^{\left( nm\right) }(\rho ,\phi )}{\delta \phi
}=-\int _{0}^{\rho }d\sigma \frac{\delta \sqrt{-g}T_{\mu
}^{\left( nm\right) \mu }(\sigma ,\phi )}{\delta \phi
}+\frac{\delta W^{\left( nm\right) }(0,\phi )}{\delta \phi }
\end{equation}
 Before dealing with the general case, we illustrate the problem in the simpler
case \( \varphi =\psi =\phi  \) which is the relevant one for SRG.

Consider the last term in (\ref{67}) in detail. We note
some restrictions on its possible form. 
In flat space \( (\rho =0) \)
the differential operator \( A^{(nm)} \) (\ref{62}) simplifies considerably:
\begin{equation}
\label{70}
A^{(nm)}=-\eta ^{\mu \nu }\, \partial _{\mu }\partial _{\nu }-E(\phi )\, \, ,\qquad E(\phi )=\eta ^{\mu \nu }\left( -\partial _{\mu }\phi \, \partial _{\nu }\phi +\partial _{\mu }\partial _{\nu }\phi \right) 
\end{equation}
We denote by \( \phi _{0} \) the background value (\ref{11}) of \( \phi  \),
\( E(\phi _{0})=E_{0} \), which in flat $D$-dimensional 
Minkowski space ($L=1$) 
becomes 
\begin{equation}
E_0=\frac {2a-1}{(2(1-a))^2} \frac{1}{u^2} \ .
\label{E0}
\end{equation}
Since Minkowski space must be quantum stable, the effective action
$W^{(nm)}(0,\phi )$ should have an extremum at (\ref{E0}) with
respect to arbitrary variations of $\phi$ and traceless variations
of the metric:
\begin{equation}
\frac {\delta W^{(nm)}(0,\phi )}{\delta\phi} \vert_{E=E_0} =0\ ,\qquad
\frac {\delta W^{(nm)}(0,\phi )}{\delta g^{\pm\pm}} \vert_{E=E_0} =0 \ .
\label{extr}
\end{equation}
By changing $D$ between $D=4$ ($a=\frac 12$) and 
$D=\infty$ ($a=1$) one can obtain infinitely many
values of the coefficient in front of $1/u^2$ in (76). Hence, (\ref{extr})
must  essentially hold for arbitrary value of that coefficient. This is
a very strong condition. Some implications will be discussed below.

To evaluate the second term in (\ref{67}), which at \( \rho =0 \)
represents a flat space contribution \( g_{\mu \nu }=\eta _{\mu \nu } \), 
for a general dilaton theory we
rewrite \( W^{(nm)}\left( 0,\varphi ,\psi \right)  \) as
\begin{equation}
\label{75}
W^{(nm)}\left( 0,\varphi ,\psi \right) =\frac{1}{4}\ln \int
\left( d\, \vec{f\,}\right) exp\left( -\int
d^{2}x\sqrt{-\eta }\vec{f}\mathbf1 _{2}\left(
A\right) \vec{f}\right) \ ,
\end{equation}
 where we have assumed that \( \varphi  \) and \( \psi  \) are 
 now independent
background fields. To compensate for double counting the degrees of freedom
from introducing the two-component real field \( \vec{f} 
\),  an additional
factor 1/2 has been introduced. In flat space the integral in the exponential
of (\ref{75}) can be expressed as
\begin{equation}
\label{76}
\int d^{2}x\sqrt{-\eta }\vec{f}\mathbf1 _{2}\left(
A\right) \vec{f}=\int d^{2}x\sqrt{-\eta
}\vec{f}DD^{\dagger }\vec{f},
\end{equation}
 where new differential operators \( D=i\gamma ^{\mu }e^{\psi }\partial _{\mu }e^{-\varphi } \)
and \( D^{\dagger }=D\left( \psi \leftrightarrow -\varphi \right)  \) in spinor
space have been introduced. Indeed, the right hand side of (\ref{76}) is equal
to
\begin{eqnarray}
\int d^{2}x\sqrt{-\eta }\left[ \vec{f}\left(
A+2\gamma ^{5}\epsilon ^{\mu \nu }e^{2\left( \psi -\varphi
\right) }\varphi ,_{\mu }\psi ,_{\nu }\right)
\vec{f}+\right.  & & \nonumber \\ \left.  + \epsilon
^{\mu \nu }e^{2\left( \psi -\varphi \right) }\varphi ,_{\mu
}\partial _{\nu }\left( \vec{f}\gamma
^{5}\vec{f}\right) \right] , & \label{77}
\end{eqnarray}
which may be used to prove (\ref{76}) after integration by parts.
Therefore, (\ref{75}) becomes
\begin{equation}
\label{78}
W^{(nm)}\left( 0,\varphi ,\psi \right) =\frac{1}{4}\ln  \det 
\left( DD^{\dagger }\right) .
\end{equation}
 For the \( \zeta  \)-function of the operator \( DD^{\dagger } \) we use
its representation in terms of an inverse Mellin transform of the heat kernel
\begin{equation}
\label{79}
\zeta _{DD^{\dagger }}(s)=\frac{1}{\Gamma (s)}\int _{0}^{\infty
}dt\, t^{s-1}{\rm Tr} \exp \left( -tDD^{\dagger }\right) .
\end{equation}
This yields the variation of \( \zeta  \) with respect to \( \varphi  \)
and \( \psi  \):
\begin{eqnarray}
\delta \zeta _{DD^{\dagger }}(s) & = & \frac{1}{\Gamma (s)}\int
_{0}^{\infty }dt\, t^{s-1}Tr\sum \frac{\left( -t\right)
^{n}}{n!}\left( 2\delta \psi \left( DD^{\dagger }\right)
^{n}-2\delta \varphi \left( D^{\dagger }D\right) ^{n}\right)
\nonumber \\
 & = & \frac{2}{\Gamma \left( s\right) }\int _{0}^{\infty
}dtt^{s}Tr\, (-2\delta \psi DD^{\dagger }\exp \left( -tDD^{\dagger
}\right) +2\delta \varphi D^{\dagger }D\, \exp \left( -tD^{\dagger
}D\right) \nonumber \\
 & = & \frac{2\Gamma \left( 1+s\right) }{\Gamma \left( s\right)
}Tr\left( -2\delta \psi D\, D^{\dagger }\left( -tD\, D^{\dagger
}\right) ^{-s-1}+2\delta \varphi D^{\dagger }D\left(
-tD^{\dagger }D\right) ^{-s-1}\right) \nonumber \\
 & = & -2sTr\left( \left( D\, D^{\dagger }\right) ^{-s}\delta
\psi -\left( D^{\dagger }D\right) ^{-s}\delta \varphi \right)
\label{80}
\end{eqnarray}

Thus the introduction of \( DD^{\dagger } \) has provided a
means to achieve multiplicative factors for the two variations -
at least in flat space, but this is sufficient for our purpose. 
By differentiating (\ref{80}) with respect to \( s \) one
arrives at
\begin{eqnarray}
\delta \zeta ^{\prime }_{DD^{\dagger }}\left( 0\right) & = &
-2\left( \zeta \left( 0\mid \delta \psi ,DD^{\dagger }\right)
-\zeta \left( 0\mid \delta \varphi ,D^{\dagger }D\right) \right)
\nonumber \\
 & = & -2\left( a_{1}\left( \delta \psi ,DD^{\dagger }\right)
-a_{1}\left( \delta \varphi ,D^{\dagger }D\right) \right)\, .
\label{81}
\end{eqnarray}
To evaluate \( a_{1} \) in the first term on the right hand side
of (\ref{81}) again the method of \cite{Gilkey} is applicable. Introducing
yet another type of differential operator in spinor space, we represent the
operator \( DD^{\dagger } \) as
\begin{eqnarray}
DD^{\dagger } & = & \left( \hat{g}^{\mu \nu }\mathcal{D}_{\mu
}\mathcal{D}_{\nu }+\mathcal{E}\right) ,\nonumber \\
\mathcal{D}_{\nu } & = & \partial _{\nu }+\psi ,_{\nu }-\varphi
,_{\nu }-\gamma ^{5}\epsilon ^{\mu }\, _{\nu }\varphi ,_{\mu }\,
,\, \, \, \hat{g}^{\mu \nu }=e^{2\left( \psi -\varphi \right)
}\eta ^{\mu \nu },\nonumber \\
\mathcal{E} & = & \hat{g}^{\mu \nu }\left( \hat{\nabla }_{\mu
}\, \hat{\nabla }_{\nu }\varphi \right) ,\label{82}
\end{eqnarray}
and again use the general result (\ref{44}).  The covariant
derivative \( \hat{\nabla }_{\mu } \) refers to the present
metric \( \hat{g}_{\mu \nu } \).  The second heat kernel
coefficient \( a_{1} \) for the operator \( D^{\dagger }D \) is
obtained by the simple replacement \( \varphi \rightarrow -\psi
,\, \psi \rightarrow -\varphi .  \)

Precisely at this point we should stress that the operator \(
DD^{\dagger } \) is only hermitian in the Dirac sense (``\(
\gamma ^{0}- \)hermitian'').  Nevertheless we see strong
arguments in favour of our treatment from our result for the
effective action (section 5).

{}From (\ref{81}) with the flat d'Alembertian (Laplacian) \( \Delta =\eta ^{\mu \nu }\partial _{\mu }\partial  \)\( _{\nu } \)
\begin{equation}
\label{83}
\delta \zeta ^{\prime }_{DD^{\dagger }}\left( 0\right) =-\frac{1}{3\pi }\int d^{2}x\sqrt{-\eta }\left[ \delta \psi \left( 2\Delta \varphi +\Delta \psi \right) +\delta \varphi \left( 2\Delta \psi +\Delta \varphi \right) \right] 
\end{equation}
 follows. We have retained the determinant also for the flat metric in order
to cover the case of light coordinates (\ref{12}) where \( \eta = \) det \( \eta \neq - \)1.
Also the covariantized version of eq.\ (\ref{83}) can be obtained then very
easily by means of the replacement \( \sqrt{-\eta }\Delta \rightarrow \sqrt{-g}\Box = \)
\( \sqrt{-g}g^{\mu \nu }\nabla _{\mu }\nabla _{\nu } \). 

 Now all variations of the effective action 
\( W^{\left( nm\right) }\left( \rho ,\phi ,\psi \right)  \)
with respect to all background fields can be summarized:
\begin{eqnarray}
\frac{\delta W^{\left( nm\right) }}{\delta \varphi } & = &
-\frac{1}{12\pi }\sqrt{-\eta }\left( 6\eta ^{\mu \nu }\partial
_{\nu }\left( \rho \partial _{\mu }\varphi \right) +2\Delta \rho
-2\Delta \psi -\Delta \varphi \right) ,\label{84} \\
\frac{\delta W^{(nm)}}{\delta \psi } & = & -\frac{1}{12\pi
}\sqrt{-\eta }\left( \Delta \rho -2\Delta \varphi -\Delta \psi
\right) ,\label{85} \\
\frac{\delta W^{\left( nm\right) }}{\delta \rho } & = &
-\frac{1}{12\pi }\sqrt{-\eta }\left( -\Delta \rho -3\eta ^{\mu
\nu }\left( \partial _{\mu }\varphi \right) \left( \partial
_{\nu }\varphi \right) +2\Delta \varphi +\Delta \psi \right)
.\label{86}
\end{eqnarray}
For arbitrary functions \( \varphi \left( \phi \right) \) and \(
\psi \left( \phi \right) \) in terms of the dilaton \( \phi \), 
eqs.\  (\ref{67}) and (\ref{68}) allow the evaluation of 
\( T_{--}^{\, \, \, \, 
(\phi )} \) in (\ref{59}) for any dilaton theory.  An
important check of our calculation is provided by the
integrability conditions \( \delta ^{2}W^{\left( nm\right)
}/\delta \varphi \delta \rho = \delta ^{2}W^{\left( nm\right)
}/\delta \rho \delta \varphi \) etc.  We take their validity as
an a posteriori argument for the validity of our derivation of
\( \delta W/\delta \phi .  \) For \( SRG \) the special choice
\( \varphi =\psi =\phi \) with (\ref{67}), (\ref{68}) yields
\begin{equation}
\label{87}
\frac{1}{\sqrt{-g}}\frac{\delta W^{SRG}}{\delta \phi
}=-\frac{1}{12\pi }\left( 6\partial ^{\mu }\left( \rho \partial
_{\mu }\phi \right) +2\Box \rho -3\Box \phi \right) .
\end{equation}

\subsection{Radiative flux}

With (90) we are now in the position to calculate the
contribution (\ref{59}) to the flux.  By simple partial
integrations we find at finite \( z \) or \( u \) {\it complete
cancellation of the integral } with the one in (\ref{58})
(except for the total divergence there):
\begin{equation}
\label{88}
T_{--}^{\left( \phi \right) }=-T_{--}^{(dil)}+\frac{1}{16\pi
}\left[ 2\left( \partial _{z}\phi \right) \left( \partial
_{z}\rho \right) +2\rho \left( \partial _{z}\phi \right)
^{2}+\left( \partial _{z}\phi \right) ^{2}-\partial _{z}^{2}\phi
\right] _{u_{h }}^{u}
\end{equation}
In the square bracket we have collected all the total derivatives. 
Eq. (\ref{88})
yields the flux at any \( u. \) At \( u\rightarrow \infty  \) the asymptotic
fluxes from \( T_{--}^{\left( \phi \right) } \)~and 
\( T_{--}^{\left( dil\right) } \)
exactly cancel because the square bracket in (\ref{88}) vanishes  
there due to the eq.\ (\ref{11}). Thus, 
the {\em flux at infinity exactly coincides with the one for 
minimal coupling} \cite{KV99/H}
\begin{equation}
\label{89}
T_{--}^{\left( nm\right) }\mid _{as}=T_{--}^{\left( min\right) }\mid _{as},
\end{equation}
and the simple thermodynamical argument, presented in that
connection \cite{Chris 77}  is also (at least superficially) applicable to the
nonminimal case as well.  It should be stressed that
the cancellation and thus (\ref{89}) { \em is independent of
the specific background}, as along as the latter yields a
vanishing contribution at \( u\rightarrow \infty \) in the last
term of (\ref{87}).  This result is highly welcome, because
the integral for \( T_{--}^{(dil)} \) can be evaluated easily,
say for \( SRG \) from arbitrary \( D>3, \)
\begin{equation}
\label{90}
T_{--}^{\left( dil\right) }=-\frac{9}{2}\, \frac{D-2}{\left(
D-1\right) }\, T_{--}^{\left( min\right) },
\end{equation}
and its negative contribution by far outweighs \( T_{--}^{\left(
min\right) } \).  Of course, for a generalized dilaton theory
with general nonminimal coupling \( \varphi \left( \phi \right)
\) in a general dilaton-dependent norm (\ref{60})  with \(\phi 
\to \psi (\phi ) \) there may be  choices of $\varphi$ and $\psi$ 
which allow \(
T_{--}^{\left( dil\right) }=0.  \) For the simplest linear
ans\"atze \( \varphi =\alpha \phi ,\, \, \psi =\beta \phi \) it
has been shown in \cite{KLV 97/H} that for certain relations
between \( \alpha \) and \( \beta \) this is indeed the case. 
But as long as there is no valid argument for such a specific
choice this had not been regarded an attractive alternative.

Now we turn to the flux at finite \( u \), and especially to the
region near the horizon \( u\approx u_{h} \).  Inserting
(\ref{11}) and (\ref{12}) into the square bracket for finite
values of \( u \) (radius) yields a flux different from \(
T_{--}^{\left( min\right) } \):
\begin{equation}
\label{91}
T_{--}^{\left( nm\right) }=T_{--}^{\left( min\right) }+
\frac{1}{16\pi }\frac{L^{2}}{u^{2}}\ln \, L
\end{equation}

We emphasize that -- as in \( T_{(--)}^{(min)} \) -- the condition
of an Unruh vacuum ``miraculously'' produced a zero
of {\em second} order at the horizon (\( L\simeq \left( u-u_{h}\right)
\)).  However, in global coordinates (\ref{32}) with factor \(
\left( u-u_{h}\right) ^{-2} \) now a residual logarithmic
divergence \cite{BF 99/2} is encountered.  For this reason a
more detailed discussion of the freedom given by renormalization
of the one-loop conformal anomaly and of the \( \phi \)-anomaly
is necessary, because this will affect the nonasymptotic region. 
The (regularized) divergent terms in the \( \zeta \)-function
regularisation in (\ref{78}) are given by \( \zeta ^{\prime
}_{A}\left( 0\right) \), and not by its variation as in 
(\ref{37}).

In principle here two normalization parameters 
\( \mu  \) and \( \mu ^{\prime } \)
are needed because two differential operators 
\( \left( A\, \, and\, \,D D^{\dagger }\right)  \)
appeared in our calculation: 
\begin{equation}
\label{92}
W^{\left( ren\right) }=\ln  \mu\, a_{1}\left( 1,\,A \right) 
+\ln \mu^{\prime}\, a_{1}\left( 1,\, DD^{\dagger }\right) 
\end{equation}

{}From (\ref{44}) it can be verified easily that the only term which
is not a total derivative, in both terms of (\ref{92}) for 
\( SRG \) \( \left( \varphi =\psi =\phi \right)  \)
becomes (with a convenient redefinition of $\mu$)
\begin{equation}
\label{93}
W_{\left( SRG\right) }^{\left( ren\right) }=
-\frac{\ln \mu }{16\pi }\int d^{2}x\sqrt{-g}\left( \nabla \phi \right) ^{2}.
\end{equation}
The contribution to \( T_{--} \) is most easily found by
interpreting (\ref{93}) as \( \int\! dz \left( \partial
_{z}\phi \right) \delta W/\delta \phi \) in the sense of the
last term of (\ref{55}), taking the conformal gauge where
(\ref{93}) is independent of \( \rho \).  Together
with (\ref{91}) the total flux at any (finite or infinite) \( u
\) becomes
\begin{equation}
\label{94}
T^{(nm,ren)}_{--}=T^{(nm)}_{--}+T^{\left( ren\right) }_{--}=
T^{\left( min\right) }_{--}+\frac{1}{16\pi }\frac{L^{2}}{u^{2}}
\ln \left( \frac{L}{\mu }\right) .
\end{equation}

Clearly no such renormalization term can arise from the flux for minimally
coupled scalars. So (\ref{94}) is the complete result for
SRG. It does not depend on the explicit form of the Killing norm \( L, \) i.e..
it is true for a general background. Fixing \( \mu  \) at an arbitrary value
\( u_{0} \) of \( u \) so that \( L \)\( \left( u_{0}\right) =\mu , \) at
\( u_{0} \) the renormalized flux coincides with the minimal one. In any case,
the logarithmic divergence at the horizon in global coordinates cannot be 
eliminated. However, that singularity is an integrable one. It cannot
lead to any ``measured'' infinity since an energy measurement should
take a finite time. The main trouble with this singular term is that
it does not seem to
appear in full $D$-dimensional theory \cite{Chris 77}.
On the other hand, the form of the logarithm in (\ref{94}) suggests a
relation to renormalization. Indeed, such terms are typical for the effective
action in a massless theory. In four-dimensional field theory a corresponding
contribution would be $L^4 \ln L$ which is harmless in the present context.
The only exception is the case when the action is defined completely by
an anomaly (e.g. the  conformal one). 
In our calculations we had to use also the 
second anomaly (the ``$\phi$-anomaly''). We may conjecture that the 
logarithmic term can be avoided if one uses the $D$-dimensional conformal
coupling (cf.\ footnote 3) 
instead of the $D$-dimensional minimal one (\ref{3}). Such a 
modification is  needed anyhow in order to make the comparison with the Christensen 
and Fulling calculations  \cite{Chris 77} more direct.

\section{Effective Action}

The three eqs.\ (\ref{84})-(\ref{86}) allow immediate functional integration
of the complete effective action \cite{KV99/H}
\begin{eqnarray}
W^{\left( nm\right) } & = & -\frac{1}{24\pi }\int \,
d^{2}x\sqrt{-\eta }\left( -\rho \Delta \rho +2\psi \Delta \rho
-\psi \Delta \psi -6\rho \left( \partial _{\mu }\varphi \right)
^{2}+\right.  \nonumber \\ 
& & \left.  + 4\varphi \Delta \rho
-4\varphi \Delta \psi -\varphi \Delta \varphi \right) \; ,
\label{95}
\end{eqnarray}
which can be written covariantly (\( \sqrt{-\eta }\Delta
\rightarrow \sqrt{-g}\Box ,\sqrt{-\eta }\Delta \rho
=-\sqrt{-g}R/2) \) as
\begin{eqnarray}
W^{\left( nm\right) } & = & -\frac{1}{24\pi }\int \,
d^{2}x\sqrt{-g}\left[ -\frac{1}{4}R\, \Box ^{-1}R+3\left( \nabla
\varphi \right) ^{2}\Box ^{-1}R-2R\left( \psi +\varphi \right)
+\right.  \nonumber \\ 
& & \left.  + \left( \nabla \psi \right)
^{2}+ \left( \nabla \varphi \right) ^{2}+4\left( \nabla ^{\mu
}\psi \right) \nabla _{\mu }\varphi \right] +W^{\left(
ren\right) }\left( \mu ^{\prime },\mu \right) \label{96}\; .
\end{eqnarray}
The first term in (\ref{96}) represents the Polyakov action
\cite{POL 81} for minimal coupling \( \varphi =\psi =0 \) of the
scalar fields.  \( \varphi \left( \phi \right) \) and \( \psi
\left( \phi \right) \) encode a general dilaton coupling of the
scalars and of the dilaton-dependent measure, respectively. 
Thus eq.\  (\ref{96}) generalizes the Polyakov action to the one for 
general non-minimal coupling to the dilaton field \cite{KV99/H}. 
The appearance of a nonlocal term should be emphasized.  A
functional integral applied to a bounded region in space time
always contains ambiguities with respect to eventual surface
variables.  In that case (\ref{96}) may acquire further (here
undetermined) contributions.  The term \( W^{\left( ren\right)
}\left( \mu ^{\prime },\mu \right) \) depending on the
renormalization points \( \mu ^{\prime },\mu \) (as in
(\ref{92})) has been discussed above for \( SRG \).  Also in the
general case (99) it only receives contributions from \( \left(
\nabla \varphi \right) ^{2} \) and \( \left( \nabla \psi \right)
^{2} \).

Again \( SRG \) from \( D \) dimensions (\( \varphi =\psi =\phi  \))
is of special interest:
\begin{eqnarray}
W^{SRG}=\frac{1}{96\pi }\int \, d^{2}x\sqrt{-g}\, \left[ R\Box
^{-1}R-12\left( \nabla \phi \right) ^{2}\Box ^{-1}R+12\phi
R-24\left( \nabla \phi \right) ^{2}\right] & & \nonumber \\
+W_{SRG}^{\left( ren\right) }\left( \mu \right) \, \, \, \, \,
\, \, \, \, \, \, \, \, \, \, \, \, \, \, \, \, \, \, \, \, \,
\, \, \, \, \, \, \, \, \, \, \, \, \, \, \, \, \, \, \, \, \,
\, \, \, \, \, \, \, \, \, \, \, & & \label{97}
\end{eqnarray}

The second, nonlocal term was not present in the analogous
formula for the full effective action in \cite{Muk 94}.  The
first three terms, however, appear in the ``uncorrected''
effective action there.  In the first ref.\  \cite{BH 97/1} all
four terms, but the last two with different factors, can be
found. Similar actions were also used in refs. 
\cite{BH/2,Mik 98}.

Usually, the full effective action including the conformally
invariant part is available as a power series in a small
parameter\cite{Vilk}.  No such parameter exists for the BH
background.  Therefore, the closed form of our action (\ref{97})
is essential.  Two previous famous examples where such a closed
form could be obtained were the Polyakov and WZNW actions.  In
those cases the effective actions were completely defined by the
corresponding anomalies.   It is remarkable that 
for general dilaton theories in \( D=2
\) we encounter  a similar situation, because the ``\( \phi
\)-anomaly'' (\ref{87}) can be interpreted as carrying the
information of part of the ``complete'' D-dimensional conformal
anomaly.

Starting with \cite{Muk 94} there have been further attempts to compute
the flux component \( T_{--} \) from an effective action like (\ref{96}),
because it is natural 
to assume that a simple functional derivative with respect
to the metric, as in the definition (\ref{20}), applied to the effective action
directly yields the desired result. However, we should recall that such a procedure
is known to be a quite delicate matter \cite{Nov 89}. Even without dilatons
(as in the theory with minimal coupling) the proper choice of the asymptotics for
the inverse d'Alembertian is very important in order to obtain correctly the
(known) final outcome, already even for minimal coupling. In the presence of
dilaton fields as in (\ref{97}) we are not aware of any analysis which shows
that there exists an appropriate choice at all. 

This uncertainty reflects a basic weakness from which, in our opinion,
all approaches are bound to suffer which use the existence of an integrated
effective action as an essential intermediate step. Such an action describes a
UV effect from quantum corrections, i.e. in coordinate space it is certainly
only correct \textit{locally.} This is also consistent with the rules for 
functional
differentiation, which in an expression like (\ref{97}) require sufficiently
strong vanishing of the fields at the infinite boundaries of the integration
in order to be able to perform partial integrations without 
surface contributions.
But the region where the flux is needed here is precisely that boundary (at
infinity)! There the flux, the functional derivative with respect 
to the metric,
should give a nonvanishing result. Also the metric itself does not 
vanish there,
but becomes Minkowskian. The main advantage of the approach used in our work
\cite{KV99/H} is that the input from the one loop quantum effects entered
\textit{locally}. The subsequent integral from the horizon to any value of \( u, \)
including infinity is a trivial, well-defined ordinary one. 

Several remarks on the validity of our method of evaluation of
the effective action are important. The part depending on the conformal
anomaly gives a well-defined local contribution to the Hawking flux
(at least in our approach).  To calculate  $W^{(nm)}(0,\phi )$ we have added
a total derivative to the classical action,  thus replacing the operator
$A$ by $DD^{\dag}$. The latter operator is not hermitian with respect
to the standard inner product for the scalar field and, therefore,
corresponds to a different path integral measure. The effective actions
calculated with $A$ and $DD^{\dag}$ differ by an ``anomaly'' 
\footnote{We are grateful to Dmitri Fursaev and Andrei Zelnikov
for informing us about their calculations demonstrating that this
``anomaly'' is indeed non-zero.}. We believe that it is the operator
$DD^{\dag}$ that should be used in calculations of $W^{(nm)}(0,\phi )$
rather than $A$ itself. Our effective action evidently satisfies the
conditions (\ref{extr}) because the only term which contributes to
the variations, $(\nabla \phi )^2$, can be removed completely by adjusting
the renormalization scale $\mu$. On the other hand, it is hard to see
how (\ref{extr}) can be satisfied by the effective action for the
operator $A=-\Delta - c/u^2$.

\section{Backscattering}

Our result (\ref{94}) undoubtedly is suggestive, but it certainly requires
further improvement. It is a central feature of the CF approach that the input
is ``minimal'' in the sense that it explores the consequences of EM-conservation
only. In the \( D=2 \) version even everything seems to be completely determined
by the anomaly and there is, at first glance, no room for the inclusion of backscattering,
an effect which in \( D=4 \) leads to complications \cite{Chris 77},\cite{Nov 89}.
They are related to the replacement of the classical free-field equation \( \partial _{+}\partial _{-}f=0 \)
for minimally coupled scalars in conformal gauge by the more complicated one
\begin{equation}
\label{98}
\left( \partial _{+}u\, \partial _{-}+\partial _{-}u\, \partial _{+}\right)
 f=0.
\end{equation}
The redefinition \( f=\widetilde{f}/\sqrt{u} \) as a consequence 
of  (\ref{60}),
i.e.\ to \( s \)-wave amplitudes, yields
\begin{eqnarray}
\label{99} &&
\partial _{+}\partial _{-}\widetilde{f}+V\, \widetilde{f}=0\, , 
\\
\label{100} &&
V=\frac{1}{2}\frac{\left( \partial _{+}u\right) \left( \partial _{-}u\right) }{u^{2}} - 
\frac{\partial _{+}\partial _{-}u}{u}\, ,
\end{eqnarray}
describing the propagation of \( s \)-wave matter in a space-dependent
``potential'' (cf.\ (\ref{11}), (\ref{12}) for a fixed BH background):
\begin{equation}
\label{101}
V=\frac{1}{4u}L^{\prime }L-\frac{1}{8u^{2}}L^{2} 
\end{equation}

In their seminal paper \cite{Chris 77} Christensen and Fulling
computed the EM conservation in \( D=4 \) for spherical
symmetry..  There more undetermined functions remained in the \(
4D \) EM-tensor when it had been restricted to spherical
symmetry.  The authors proposed to use the \( 4D \) conformal
anomaly \textit{and} an estimate of backscattering effects in
order to obtain information about those undetermined parts.  In
contrast, in the present \( SRG \) approach everything seems to
be fixed by the anomaly alone as long as (cf.  Section 4.1) the
assumption is made that the classical background of the scalar
field vanishes (\( f_{class}=0 \)).

For minimally coupled scalars this assumption was consistent with
the decoupling of \( f_{class} \) from the problem of outgoing radiation: There
a BH forms ``immediately'' from the incoming flux \cite{Chop}, and the (classical)
BH remains stable (\( T_{--}^{(class)}=0 \) trivially follows from the condition
\( T^{(class)}_{--}\mid _{u_{h}}=0 \)).

In the present case the period of the formation of the BH cannot
be separated as cleanly from the situation after that.  There
will be still an outgoing flux of matter (as confirmed also
by numerical computation \cite{Chop}).  Thus a solution for \(
f_{class}\neq 0 \) to the coupled (classical) system of e.o.m.-s
following from (\ref{2}) together with (\ref{4}) would be
needed.  The matter equation (in conformal gauge) would be
(\ref{99}).  On that \( f_{class} \) the quantum corrections are
``riding''.  A more transparent (approximate) view of the
situation is as follows: The flux of matter will influence the
space curvature through the classical Einstein equation
especially near the horizon by being backscattered from a
maximum of some effective potential related to \( V \) in
(\ref{100}), which for \( SRG \) at \( D=4 \) (with traceless
EM-tensor, cf.  footnote 3) is situated at \( u\approx
\frac{3}{2}u_{h}=3M_{ADM} \) \cite{Chris 77}.  At infinity its
contribution to \( T_{--}^{\left( nm\right) }\mid _{as} \) is
negligible, but in the region \( u_{h}\leq u\simeq 2u_{h} \) it
changes the geometric background to be used in the integration
of the ``EM-nonconservation'', eq.\  (\ref{53}).  As a consequence,
the flux at infinity will be attenuated by a certain amount
(``grey factor'').  We refer the reader to the extensive
literature on this subject (e.g.\  \cite{Nov 89,Chop}).

The conclusion to be drawn for our present strictly \( 2D\,
(SRG) \) approach is that all these additional considerations
can be (and should be) included as well in a purely \( 2D \)
approach.  We do not see any obstacle for adapting the \( 4D \)
arguments to the \( SRG \) setting directly.  Of course,
features specific for \( 4D \) and related to the additional
angular dependence will always remain outside the scope of a \(
2D \) calculation.

\section{Comparison with other approaches, outlook}

The necessity to modify the energy-momentum tensor for non-minimally
coupled scalars in order to avoid negative flux was recognized already
in  \cite{Muk 94}. To this end attempts have been made
\cite{Muk 94,LOM 98}
to calculate the scale invariant non-local part of the effective
action. Due to the ambiguities in definition of the inverse
Laplacian this task is technically quite complicated. Usually
such calculations are valid in a narrow region of the
parameter space only, reflecting the weakness of the effective action
approach. The effective action of \cite{Muk 94} becomes 
singular in the limit of  Minkowski space while the energy-momentum tensor
of \cite{LOM 98} fails to satisfy the conservation condition
(\ref{53}).

The importance of the modified conservation law (\ref{53}) was noted in
\cite{BF 99} where some modifications of the energy-momentum
tensor were suggested. 

The authors of the second ref.\ \cite{Mik 98} proposed the 
introduction of  two auxiliary fields, $\chi$ and $\psi$, 
\begin{equation}
\Box\chi =(\nabla\phi )^2 \ ,\qquad \Box \psi =R\; ,
\label{auxil}
\end{equation}
to avoid non-locality in the effective action. Arbitrary constants
arising from solving the equations (\ref{auxil}) for $\chi$ and
$\psi$ were used later by Balbinot and Fabbri \cite{BF 99/2}
to model different vacuum states. In particular, they obtained that
the conformal anomaly induced effective action either produces
negative Hawking flux or gives logarithmic behaviour at the horizon
as in our eq.\ (\ref{91}). We can even further sharpen this result.
Indeed, since the fields $\chi$ and $\psi$ were used to represent
the non-local term $\int (\nabla \phi)^2 (1/\Box )R$ in two
different ways, a consistency condition must hold
\begin{equation}
\int d^2x \sqrt{-g} (\nabla \phi )^2 \psi =
\int d^2x \sqrt{-g} R \chi \ .
\label{CC}
\end{equation}
This condition turns out to be very restrictive. For example,
it immediately gives $C=0$ in the notations of \cite{BF 99/2}.
{}From eq.\ (13) of \cite{BF 99/2} it is seen that it is not
possible to remove the logarithmic term from the energy-momentum
tensor. 

Our approach has successfully passed various consistency tests
and gives physically plausible results. Namely, the Hawking
flux is positive and the Hawking temperature is defined by
the surface gravity in the usual way. However, we should mention
certain weak points of this approach as well. The first problem is
common for nearly all approaches dealing with non-minimal
coupling in 2D. This is the presence of the $L^2 \ln L$ term
in the energy momentum tensor near the horizon. In our view
this singularity is quite mild and may be an artefact of the 2D
renormalization scheme. Possibly it can be removed by using
the $D$-dimensional conformal action for scalars. On the other 
hand, we are not aware of any $D = 4$ calculation which 
explicitly forbids such a term. A more serious
problem for us is to justify the transition to the operator 
$DD^\dag$ in the 
calculation of $W^{(nm)}(0,\phi )$. Even though we have 
presented certain arguments in favour of this choice,  as compared
to direct calculations with the operator $A$, the whole procedure
still involves a step which lacks complete mathematical
rigour.

In the present paper we have critically reviewed various
approaches to Hawking radiation from 2D black holes and
presented an approach which, in our opinion, so far is the most
appropriate one for a proper description of this process, 
although we 
must admit that this problem cannot be considered to be settled
as yet.  To clarify the situation a direct calculation starting
from 4D {\it quantum} theory with 4D renormalization conditions
is needed urgently.

Many questions of black hole physics are outside the
scope of the present review. For more information an interested
reader should 
 consult the large literature on the subject from which
we recommend for a recent collection ref.\ \cite{More}.

\noindent \textbf{\Large Acknowledgements}{\Large \par}

\noindent The authors are grateful for valuable discussions 
and/or correspondence
with R. Balbinot, M. Bordag, G. Cognola,
A. Fabbri, D. Fursaev, J. Louko, V. Mukhanov, A. Wipf and A. Zelnikov.
We thank F. Hehl for his kind invitation to write this review.
This work has been supported
by Fonds zur F\"{o}rderung der wissenschaftlichen Forschung 
(Austrian Science Foundation), Project P 12.815-TPH, the Alexander
von Humboldt Foundation and RFBR, grant 97-01-01186.

\end{document}